Systems biology

# A nonparametric significance test for sampled networks

Andrew Elliott[1,*], Elizabeth Leicht[1], Alan Whitmore[2], Gesine Reinert[3] and Felix Reed-Tsochas[1,4,*]

[1]CABDyN Complexity Centre, Saïd Business School, University of Oxford, Oxford OX1 1HP, UK, [2]e-Therapeutics plc, Long Hanborough, OX29 8LN, UK, [3]Department of Statistics, University of Oxford, Oxford, UK and [4]Oxford Martin School, University of Oxford, Oxford, UK

*To whom correspondence should be addressed.
Associate Editor: Alfonso Valencia

## Abstract

**Motivation:** Our work is motivated by an interest in constructing a protein–protein interaction network that captures key features associated with Parkinson's disease. While there is an abundance of subnetwork construction methods available, it is often far from obvious which subnetwork is the most suitable starting point for further investigation.
**Results:** We provide a method to assess whether a subnetwork constructed from a seed list (a list of nodes known to be important in the area of interest) differs significantly from a randomly generated subnetwork. The proposed method uses a Monte Carlo approach. As different seed lists can give rise to the same subnetwork, we control for redundancy by constructing a minimal seed list as the starting point for the significance test. The null model is based on random seed lists of the same length as a minimum seed list that generates the subnetwork; in this random seed list the nodes have (approximately) the same degree distribution as the nodes in the minimum seed list. We use this null model to select subnetworks which deviate significantly from random on an appropriate set of statistics and might capture useful information for a real world protein–protein interaction network.
**Availability and implementation:** The software used in this paper are available for download at https://sites.google.com/site/elliottande/. The software is written in Python and uses the NetworkX library.
**Contact:** ande.elliott@gmail.com or felix.reed-tsochas@sbs.ox.ac.uk
**Supplementary information:** Supplementary data are available at *Bioinformatics* online.

## 1 Introduction

Network sampling is used in many different fields, such as biology (Lim *et al.*, 2006) and sociology (Bernard *et al.*, 2010; Frank and Snijders, 1994). Many studies sample a known network to produce a subnetwork which is believed to be more relevant to their research goals than the initial network such as a subnetwork associated with metabolism. Frequently protein–protein interaction (PPI) networks are sampled to form subnetworks that are associated with the disease or cellular processes of interest e.g. Hwang *et al.* (2008); Lim *et al.* (2006); Gao *et al.* (2011); Goehler *et al.* (2004); Chuang *et al.* (2007); Sharma *et al.* (2015); Ghiassian *et al.* (2015). An advantage of such sampling is that on a small network an in-depth analysis, such as verifying existing links, may be feasible. Network sampling can also reflect empirical limitations such as the availability of partial data for a given network (Bernard *et al.*, 2010; Frank and Snijders, 1994), or the exclusion of vertices that cannot be detected (Salganik, 2006), with consequences for measured network statistics (Kossinets *et al.*, 2006).

Subnetwork construction methods are not without their own problems, since they may induce artefacts in the subnetworks that they generate. Even the use of a PPI interactome as a starting point already intrinsically reflects the effects of sampling, since different experimental methods vary in their ability to identify particular interactions. There are also inherent biases in the levels of evidence available for different interactions, and generally PPI networks are known to exhibit high rates of both false positives and false negatives (Ali *et al.*, 2011). Notably there is no gold standard method for constructing a network representing a cellular process, although several techniques have been proposed to achieve this aim. Some studies test interactions experimentally between a subset of proteins believed to be important to a disease process (Goehler *et al.*, 2004; Lim *et al.*, 2006). Another approach is to locate proteins present in the same cellular compartment as the process of interest, and add edges between these proteins using a PPI database (Gao *et al.*, 2011). One can also form subnetworks from a larger PPI dataset using a seed list in conjunction with a construction method, e.g. snowball sampling (Martin *et al.*, 2010), path based methods (Berger *et al.*, 2007), Steiner trees (White and Ma'ayan, 2007) or the inclusion of nodes based on significance testing (Ghiassian *et al.*, 2015; Sharma *et al.*, 2015). Finally one can also take a network directly from a pathway database e.g. KEGG (Hwang *et al.*, 2008).

The sampling techniques in this paper start from a list of seed nodes and apply what we call a *construction method* to generate a subnetwork from these seed nodes. Seed nodes are typically believed to have certain attributes or associations, e.g. proteins implicated in a disease. As the underlying PPI network is available, this approach is subtly different from the standard use of these network sampling techniques, namely sampling a large unknown network with the aim of estimating global properties (Bernard *et al.*, 2010; Frank, 1977; Newman, 2010). The construction methods used in this paper following prior work on biological systems (Berger *et al.*, 2007; Li *et al.*, 2012; Martin *et al.*, 2010; Shi *et al.*, 2014) are as follows: (i) snowball sampling; (ii) all paths up to a given length; (iii) all shortest paths between seed nodes. Snowball sampling has been used in biological systems through easy to use plug-ins to popular software applications; for instance the Cytoscape plug-in Bisogenet (Martin *et al.*, 2010) and to find hidden populations (e.g. drug users) in Sociology (Bernard *et al.*, 2010; Frank, 1977; Salganik, 2006). A method using all paths up to a given length has been used in biology through the Genes2Networks web app (Berger *et al.*, 2007). We are not aware of a published software package that uses shortest-path sampling, although Li *et al.* (2012) have used shortest-path sampling in a study on colorectal cancer and Keane *et al.* (2015) have used shortest-path sampling to study Parkinson's disease. It is important to note that in general the effect of network sampling on network statistics is non-trivial, and only well understood for very limited combinations of sampling methods and underlying networks. For instance, it has been shown that the degree distribution of a network uniformly sampled from a scale free network is not itself scale free (Stumpf *et al.*, 2005). To select good subnetworks, guidance about typical samples, or indeed atypical subnetworks, is required.

Here we provide a method to assess when a given subnetwork differs significantly from randomly generated subnetworks. A subnetwork which differs significantly from a random network could be viewed as containing relevant information, assuming that the comparison with the random network is meaningful. Hence a key question concerns the rules for constructing an appropriate null model, or a correctly randomized subnetwork.

As there is no generally accepted parametric model of PPI networks (Rito *et al.*, 2010), we are unable to construct a general null model based on an ensemble of PPI networks. Instead, our method compares a statistic of interest against that obtained for an ensemble of subnetworks generated from the same underlying network using a set of seed lists which are randomly chosen under certain constraints. First, we match the degree of the seeds with those of the original seed list. By contrast, the popular configuration model would match the degree of all nodes in the subnetwork. Second, there is a further feature in our null model, which relates to redundancy in the seed list. Many different seed lists may give rise to the same subnetwork. Hence given the construction method, we must also control for the construction of the seed list. Some of these seed lists can be constructed by removing nodes from the original seed list so that the subnetwork that is generated from the modified seed list is identical to the original one. We refer to the seed nodes that can be removed without changing the subnetwork as 'redundant seed nodes'. On this basis we can then construct a meaningful null model using subnetworks generated at random with the same (approximate) degree distribution as the smallest subset of the original seed list which generates the same network (the minimum seed list). We use this null model in a nonparametric significance test for features of sampled networks. Our null model allows us to assess the significance of network features given a construction method, rather than given a construction method and a fixed seed list.

The test is first illustrated using simulated stochastic blockmodel data for a network with two groups. A stochastic block model assigns each node to a group and then places edges between a pair of nodes with a fixed probability based on the group to which the node has been assigned. We demonstrate that significance under our test is correlated in all but one case with two well-known measures: accuracy (a measure of the completeness of sample) and purity (a measure of the ability of the sampling method to select nodes from the correct group). However, we note that one of the correlations is weak. We then compare subnetworks generated by two seed lists related to Parkinson's disease (PD), namely gene data from the OMIM database (Hamosh *et al.*, 2005) and a seed list derived from expression data of a PD cell model (Conn *et al.*, 2003). We find that the networks generated from the expression data seed list under the 'all shortest paths between seed nodes' sampling scheme and under the 'all paths up to length 2 (including paths of length 2)' sampling scheme have significant results under our null model (although the latter is only partially robust to parameter choice), and therefore may have interesting properties for further analysis for our work on PD.

We demonstrate the effect of redundant seed nodes, first through simulations with randomly selected seed lists. Second, we investigate the effect in our two seed lists related to PD, finding that redundant seed nodes can have a strong impact on the perceived significance of network statistics. We also demonstrate that our method compares favourably to the configuration model on this class of network sampling problems.

## 2 Materials and methods

### 2.1 Network sampling

The methods presented in this paper focus on techniques to form subnetworks using a given seed list, where we use the following three sampling techniques:

1. **Snowball Sampling** includes all nodes that are less than a given graph distance from the nodes in a seed list; an example can be found in Figure 1A. Depending on the implementation, the subnetwork can include only edges that were involved in the sampling process, or also include additional edges between sampled nodes, which we call cross-edges. In this paper we write Snow1

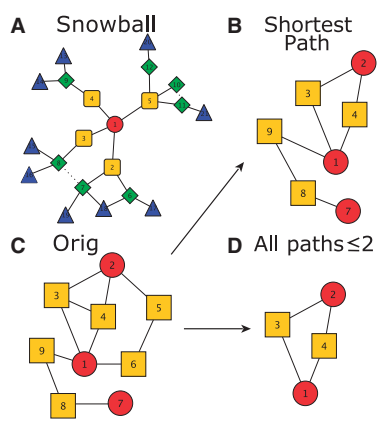

**Fig. 1.** (**A**) 2-hop Snowball Sampling Example. The seed list consists of node 1 (circle) only. The shape of the other nodes represent the distances from the seed node: squares represent nodes 1 hop from the seed, diamonds 2 hops from a seed and triangles 3 hops from a seed. Dashed edges represent cross-edges in a 2-hop snowball sample. (**B**)-(**D**) demonstrate sampling techniques based on paths. The network in (**C**) represents the unsampled network. (**B**) and (**D**) show the network in (**C**) sampled with the 'All Shortest Paths' (**B**) and Path2 (**D**) methods respectively. Seed nodes are represented by circles and other nodes are represented by squares (Color version of this figure is available at Bioinformatics online.)

for 1-hop snowball sampling, and Snow2 for 2-hop snowball sampling.

2. **All Paths $\leq$ k** (abbreviated Path2, Path3, Path4) includes all nodes and edges that are on a path between seed nodes that is less than or equal to $k$ in length. An example can be found in Figure 1D.

3. **Shortest-path Sampling** (abbreviated Shortest) includes all shortest paths between all pairs of seed proteins. An illustration can be found in Figure 1B.

To illustrate the method, and following the approach of Ratmann *et al.* (2009), we use a basket of commonly used network summary statistics, namely assortativity, average degree, clustering coefficient and number of nodes, using the following definitions. Other approaches are also available, see for example Thorne and Stumpf (2012).

- **Assortativity:** The Pearson's correlation coefficient between the degree of nodes on either side of an edge;
- **Average Degree:** The mean number of edges per node;
- **Average local Clustering Coefficient:** The average of the local clustering coefficient of each node. The local clustering coefficient is defined as the number of triangles a node is involved in divided by the number of possible triangles (i.e. the number of pairs of edges that a node has).
- **Number of Nodes:** The number of nodes in the sampled network.

We choose these summary statistics, as they are commonly used and have low computational complexity. In the case that assortativity is not defined, for example because in the Path $\leq k$ sampling method there are no paths $\leq k$ between seed nodes, or because all nodes in the sampled network have the same degree, the value of assortativity is set to 0. Similarly, when there are no possible triangles (i.e. no nodes with degree greater than 1), the clustering coefficient is set to 0.

For Path$\leq$k sampling, when a seed node is not connected to any of the other seed nodes with a path of length less or equal to $k$, this seed node is ignored for the calculation of the summary statistics. This choice is made to help interpret comparisons between subnetworks generated by different seed lists.

### 2.2 Network data
To create our PPI network we use the yeast 2 hybrid (Y2H) experimental results in the BioGRID database version 3.4.127 (Chatraryamontri *et al.*, 2013; Stark *et al.*, 2006). We remove all interactions that do not include a human protein. We then reduce the Y2H BioGRID network to its largest connected component, i.e. the graph formed from the largest group of nodes for which there is a path between any pair of nodes. The resulting network has 8292 nodes, 25 062 edges, a density of 0.00073, and an average local clustering coefficient of 0.045.

### 2.3 Seed lists
We compare two different seed lists for PD. For the first seed list, which we abbreviate *OMIM*, we assemble a list of genes known to be involved in the disease taken from the OMIM database (Hamosh *et al.*, 2005). We convert the genes to proteins in the BioGRID database using the relations provided in the BioGRID database (Chatraryamontri *et al.*, 2013; Stark *et al.*, 2006), resulting in a seed list with 16 proteins, of which 13 are present in our network, which form the OMIM seed list.

We construct the second seed list from differential expression data of 1185 genes in SH-SY5Y cells (a human cell line) before and after treatment with MPP+ (a toxin used as a model for PD) (Conn *et al.*, 2003). In Conn *et al.* (2003) 313 genes were differentially expressed, 48 of which were deemed to be statistically significant. This list includes genes that are up and down regulated by the cell when presented with MPP+. We convert the 48 significant genes to BioGRID gene identifiers. There are multiple mappings for some of these genes, resulting in 54 proteins, of which 46 are present in our network and these form the Expression seed list.

There is no overlap between the Expression seed list and the OMIM seed list. More information on the seed lists can be found in Supplementary Information S1.

### 2.4 Minimum seed lists and redundant seed nodes
We define a 'redundant seed node' as a node in a seed list that can be removed without changing the resulting subnetwork. For a given seed list we then define the (set of) minimum seed list(s) as the smallest non redundant subset (or set of subsets) of the original seed list which produces the same subnetwork.

As an example consider a triangle with nodes 1, 2 and 3. In 1-hop snowball sampling with a seed list consisting of all nodes, the set of minimum seed lists is $\{\{1\}, \{2\}, \{3\}\}$. In contrast, the set of minimum seed lists using the seed list $\{1, 2\}$ would be $\{\{1\}, \{2\}\}$. Note that $\{3\}$ is not present, as it is not a subset of the original seed list.

Computing the minimum seed list for a given subnetwork by considering all possible seed lists is computationally prohibitive. If we can guarantee that removing seed nodes does not add any previously unseen nodes or edges to the subnetwork (which all tested techniques in this paper satisfy), we can use the procedure below:

1. Remove each seed in turn and check if the number of nodes and edges in the subnetwork do not change. If not, then add the node to the list of redundant seeds.
2. Form a list of the remaining seeds.
3. Define a dummy variable L and set $L = 0$
4. For lists of redundant seeds of length $L$
5. Test if sampling with the list of the remaining seeds and the selected redundant seeds produce the same network.
6. Store all seed lists which pass the test.

7. If there are no seed lists which pass the test, set $L \rightarrow L+1$ and go to step 4.
8. Return the smallest seed list(s) that produce the same network.

However, it should be noted that it is possible that there is a smaller seed list that generates the same network that is not a subset of the original seed list. However, an advantage of our technique is that it is globally applicable and computationally tractable.

If the above procedure proves to be computationally prohibitive (which was not the case for results presented in this paper), we may be able to convert the problem to an NP-hard problem and then use current best known algorithms. For example, in the case of snowball sampling the problem of finding the minimum seed list can be converted to set cover. For further explanation and some other optimizations for this problem see Supplementary Information S3.

### 2.5 The null model

For significance testing we would ideally want to use a parametric null model (in this case a parametrized random network ensemble), but currently no suitable parametric null model exists - for discussions on PPI networks see e.g. Ali *et al.* (2011) or Rito *et al.* (2010). As an alternative we create a null model using an ensemble of subnetworks that have been generated from a random seed list of the same size and (approximate) degree sequence as the original seed list. To adjust for redundancy, we use the smallest possible seed list that generates the same subnetwork. This model replicates the effect of the sampling procedure on the network.

It is difficult to construct analytic results, due to the dependence between seed nodes, while we can construct some results see Section 3.1, it is not computationally feasible to apply them to large networks. Thus we use Monte Carlo methods.

We create a null model by estimating the underlying distribution using an ensemble of networks sampled using random seed lists of the same length and (approximate) degree distribution as the minimum seed list. We then calculate the $P$-values for the statistic of interest using a Monte Carlo test. Hence our procedure is as follows:

1. Construct the minimum seed list.
2. Generate many random seed lists with the same length and (approximate) degree distribution as the minimum seed list.
3. Generate a subnetwork for each seed list using the construction method under consideration (as described above).
4. Calculate the test statistic on each of the subnetworks.
5. Compute the $P$-value by counting the number of subnetworks with at least as extreme a test statistic as the subnetwork in question.

A $P$-value is defined as the probability, under the null model, of getting a value as least as extreme as the observed value. If $T(X)$ is a test statistic and we observe $T_{obs}$, then $p(T_{obs}) = P(T(X) \leq T_{obs})$.

Strictly enforcing the degree distribution may introduce problems in finite networks as there is a finite number of nodes of a given degree, possibly leading to a small number of random choices for some seed nodes. In order to alleviate this bias, the nodes are binned by degree from the left, such that each bin contains at least a predefined number of nodes, and the random selection of nodes is performed inside each bin. Where feasible stability testing is then performed over different bin sizes (5, 10, 20, 30 and 50) to guarantee that the result is robust to the bin size. Here we show results for bin size 20 only. Results for other bin sizes are in the Supplementary Information S5; the conclusions drawn in this paper are robust to the bin size unless otherwise stated. This is why our method for constructing the null model specifies the (approximate) degree distribution and not the exact degree distribution.

### 2.6 Benchmarking the approach

To gauge where it is appropriate to use this method, we test when the method successfully selects subnetworks that better represent the network of interest on a simple benchmark network. We construct the benchmark network with known groups from a stochastic blockmodel and then sample from it using a randomly selected seed list. We start with 4000 nodes, and assign the first 2000 nodes to Block 1 and the second 2000 nodes to Block 2. We place an edge between every pair of nodes in the same block with probability $p = 0.01$, and we place an edge between every pair of nodes in the different blocks with probability $q = 0.001$. We select 20 nodes from Block 1 to form the seed list. We sample a network using this seed list and the construction methods of interest; we record the $P$-value under the null model proposed in this paper.

We then measure the success of the sampling by looking at the *accuracy* (a measure of the completeness of the sample) and the *purity* (also called precision, a measure of the ability of the sample to select nodes from the correct block) of the classification which would classify all nodes in the sampled subnetwork as Block 1 nodes. We define $C_1$ as the set of nodes that are selected in the sample and $C_2$ is the set of nodes that have not been selected.

In this context we define accuracy $A_{cc}$ as:

$$A_{cc} = \frac{|C_1 \cap \text{Block 1}| + |C_2 \cap \text{Block 2}|}{|C_1| + |C_2|}$$

and purity $P_{ur}$ as:

$$P_{ur} = \frac{|C_1 \cap \text{Block 1}|}{|C_1|}.$$

Due to computational demands of comparing these experiments over the ensemble, we restrict the minimum bin size to 20. Here we compare seed lists for a fixed method. For an exploratory analysis we can also compare different methods for a fixed seed list.

### 2.7 Assessing a null model fit

To evaluate the significance of any statistic with respect to the possibility of it being generated by random chance, the result must be compared against a credible null model.

One basic test of the applicability of a null model to a particular random process is to test if the distribution of $P$-values of randomly generated results is uniform provided that the null distribution is continuous. We can assess this hypothesis using the following procedure:

1. Create random seed lists from a given network.
2. For each seed list create a subnetwork with the given technique.
3. Measure the statistic of interest on the subnetwork.
4. Use the null model of choice to calculate the $P$-value for the statistic of interest.

If $T_{obs}$ is drawn uniformly at random from the distribution of $T_{obs}$ and if this distribution is continuous, then under the null hypothesis the random variable $p(T_{obs})$ is uniformly distributed on [0,1]. We can therefore assess the appropriateness of the null model by performing a $\chi^2$ goodness of fit test on the distribution of $P$.

# 3 Results

## 3.1 Analytic null model statistics

The interdependence between seed nodes severely limits the range of sampling techniques and statistics for which we can define tractable analytic expressions for the distribution of the statistic of interest over an ensemble. However, one case where we can derive an analytic solution is the number of nodes in $n$-hop snowball sampling with a seed list of size $s$. Inspired by Frank (1977), we can derive the mean and variance of the number of nodes $X$ in a sampled network for a random seed list $S$ (see Supplementary Information S2):

$$E(X \mid |S| = s) = |V| - \sum_{\substack{J \subset V \\ |J|=1}} h(J,s),$$

$$\text{Var}(X \mid |S| = s) = \left( \sum_{\alpha=1}^{2} \alpha \sum_{\substack{J \subset V \\ |J|=\alpha}} h(J,s) \right) - \left( \sum_{\substack{J \subset V \\ |J|=1}} h(J,s) \right)^2, \quad (1)$$

where $|S|$ is the length of the seed list, $V$ is the set of all nodes (of size $|V|$), and $h(J, s)$ is the probability that none of the $s$ randomly chosen seed nodes are within $n$ hops of the nodes in $J$. The probability $h(J, s)$ is calculated via a hypergeometric distribution, considering the network as fixed. This approach can be extended to (approximate) degree distribution on the seed list by modifying $h$ and placing additional constraints on $S$ (see Supplementary Information S2).

The effect of seed list size on the distribution of the number of nodes in a 1-hop snowball sample in the BioGRID PPI network (Chatraryamontri *et al.*, 2013; Stark *et al.*, 2006) can be found in Figure 1 in the Supplementary Information. A small change in the number of seed nodes can have a large impact on the expected size of the network.

## 3.2 Evaluation on the benchmark data

To test whether there is a negative relationship between the $P$-value of our test and accuracy or purity, we use Kendall's $\tau$ statistic which is a measure for association. The value is in $[-1, 1]$; the closer to $\pm 1$; the stronger the association. We measure Kendall's $\tau$ with respect to the minimum of the $P$-values of the two tails, as we do not specify in which direction the statistics differ. Each of the $P$-values are computed using 10 000 Monte Carlo realizations.

The results in Table 1 show that for all of the single construction methods except for the Shortest Path construction method, there is the desired negative association (the smaller the $P$-value, the better the assignment to the block). Although, in the case of the Path2

**Table 1.** Kendall's $\tau$ for the relationship between test $P$-value and accuracy or purity in the benchmark dataset

| Method | Kendall's $\tau$ for accuracy | Kendall's $\tau$ for purity |
|---|---|---|
| Snow1 | −0.231 | −0.184 |
| Snow2 | −0.115 | −0.113 |
| Shortest | 0.534 | −0.222 |
| Path2 | −0.594 | −0.022 |
| Path3 | −0.265 | −0.173 |
| Path4 | −0.113 | −0.112 |

*Note*: The benchmark is a stochastic block model, consisting of two blocks of size 2000 with an internal connection probability of 0.01 and an external edge probability of 0.001. Further, details of this benchmark can be found in Section 2.6.

method the correlation with purity is small, however the correlation with accuracy is much stronger.

Note, here we compare seed lists for a fixed method. As part of an exploratory analysis, we can also compare different methods for a fixed seed list. We also note that the results obtained here are not independent of the parameter choices.

For differentiating between subnetwork construction methods we also investigated the trade-off between accuracy and purity. The results in Figure 2 show that care must be taken in selecting the correct construction method for the problem at hand by considering the trade-off between purity and accuracy of each of the methods. The Path4 method has can achieve high accuracy but does not achieve high purity, while Path2 achieves the highest purity overall, but has low accuracy.

## 3.3 Comparing sampling methods and seed lists for PD

When trying to construct a subnetwork which reflects a disease process, one is faced with a plethora of choices. In order to address this problem in our work on Parkinson's disease (PD) we compare how far the subnetworks deviate from random according to the null model described earlier in this paper. While we do not know if the subnetwork that deviates the most from random will contain more (or less) biological information than other subnetworks, it is possible that there are certain subsets of the sampling techniques described above that identify interesting structural features which may also be biologically meaningful. As we cannot test all possible summary statistics, we use the statistics described in Section 2.1 as a comparison.

To illustrate our approach we compare our two different seed lists for PD, OMIM and Expression (see Section 2.3 for details), across all of our sampling techniques and a reasonable parameter range.

To contrast the different sampling techniques, we compute the significance of all of the statistics in our set and select the smallest $P$-value. We test in both tails, at significance level 0.025, and as we compare 4 statistics we apply a Bonferroni correction resulting in a significance test at level $0.025/4 = 0.00625$. The significance results for the OMIM seed list and the Expression seed list can be found in

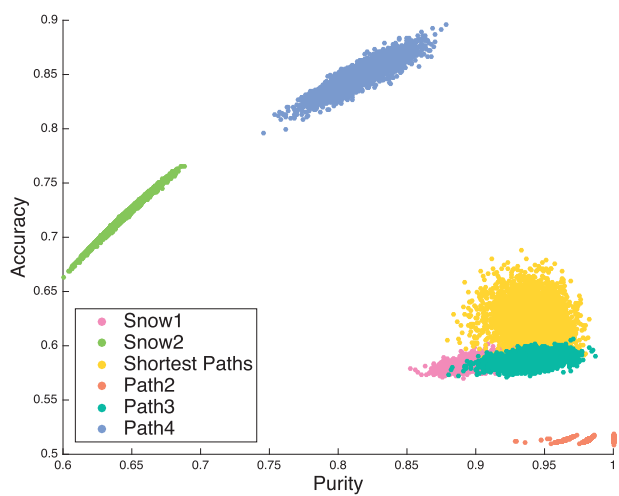

**Fig. 2.** A scatter diagram of accuracy versus purity of benchmark networks in which the sample is significant under our test (significance level $0.025/4$ due to a two-tailed adjustment and a Bonferroni correction) where colour represents the construction method used. An ideal method would have accuracy = 1 and purity = 1

Figure 3. The *P*-values are plotted on a log-scale: the higher the box, the smaller the *P*-value. The 0.00625 threshold is marked by a red line.

Two networks show promising deviation from randomly sampled networks. In the Expression seed list Path2 and Shortest Path are significant, the Path2 method is robust in all but one bin size (50) and the shortest path is robust in all bin sizes. In the OMIM seed list while Snow1 is not significant it is approaching significance with a *P*-value of 0.0063 and as such may deserve further consideration. While we cannot claim that the other networks do not have any information about the disease condition, the significance of these networks suggests that these may be a good networks on which to focus in depth analysis.

We also explored how many networks in each sampling technique have assortativity values which are assigned a value of 0. Most construction methods very rarely experience this, however 11% of the OMIM Path2 Monte Carlo test null network ensemble and 27% of the OMIM Path2 minimum Monte Carlo test null network ensemble have assortativity values that are set to 0. This is mostly due to the short seed list.

In view of Figure 2 which shows poor accuracy for Path2 sampling, our preferred subnetworks are the networks created from the Expression seed list via all shortest paths and the OMIM seed list via snowball 1. This subnetwork contains 1383 nodes of the 8292 nodes in BioGRID; it contains 4252 of the 25 062 edges in BioGRID. Its density of 0.0044 is markedly higher than the BioGRID density (0.00073), while the average local clustering coefficients are similar (0.044 versus 0.021).

### 3.4 Redundant seed nodes in PPI networks

As our null model starts with random seed lists of the same length and (approximate) degree distribution as the chosen minimum seed list, our test relies crucially on a minimum seed list. Without reducing the original seed list to a minimum seed list, the test results could be very different – we call these resulting *P*-values *perceived P-values*, the *P*-values which we would perceive if we were not to correct for redundant seeds.

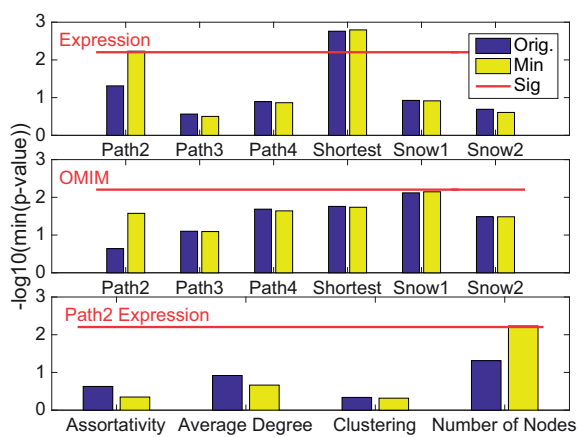

**Fig. 3.** Test results for different seed lists: smallest *P*-value, on a negative log scale. Results are shown for the Expression seed list (first panel); OMIM seed list (middle panel); and a breakdown of the *P*-value for the 4 statistics evaluated for the Path2 Expression network (final panel). Blue (left bar): original seed list; yellow (right bar): minimum seed list; red (horizontal line): significance level (0.025/4). Note due to the negative log scale on the y axis, values above the red line are significant. Each of the *P*-values are computed using 15 000 Monte Carlo realizations (Color version of this figure is available at Bioinformatics online.)

To demonstrate the effect of redundant seed proteins on perceived *P*-value of network features first we add redundant seed nodes to randomly selected seed lists in the BioGRID PPI network, and second we compare the perceived *P*-value on the networks based on PD seed lists. We illustrate our results for assortativity, average degree, clustering and number of nodes.

We investigate the importance of accounting for redundant seed proteins by comparing the significance of two seed lists that generate the same network. We construct an ensemble of random seed lists of length 25 sampled uniformly at random from all possible seed lists. For each seed list, we construct the longest seed list that generates the same network. We compute the difference between the perceived left *P*-value in the original seed list and the left *P*-value of the longest seed list. If there is little difference we would expect the results to be close to 0. The algorithm used to construct the longest possible seed list can be found in Supplementary Information S3. For simplicity in cases where there is more than one longest seed list we select one randomly.

On the BioGRID PPI network with the Snow2 construction method (Fig. 4), we observe a large difference in *P*-values in all statistics. Figure 3 shows that while adjusting for minimum seeds often does not make a large difference to perceived *P*-value, in the case where it does (Fig. 3 Expression seed list Path2), the change can be large.

Also adding redundant seed nodes to seed lists in the other sampling techniques, may result in considerable changes in *P*-value, see Supplementary Information S4. Thus, the finding that redundant seed nodes can influence the *P*-value of statistics is not restricted to our real-world examples.

### 3.5 Comparison with the configuration model

A popular null model in network science is the configuration model, which has been widely used as a null model across application domains. In the configuration model, the network is rewired randomly while preserving the degree distribution of the network (Newman *et al.*, 2001). By contrast, the configuration model does not preserve the structure induced by sampling in a network.

We compare the distribution of *P*-values for this null model and the configuration model using the method presented in Section 2.7, using 1000 randomly chosen seed lists of length 25 for assortativity and clustering on the BioGRID network (Fig. 5). Assortativity and clustering display a distribution which is approximately continuous. In the configuration model, the *P*-values under the $\chi^2$ test of assortativity and clustering are numerically equal to 0, providing strong evidence to reject the configuration model. In contrast, under our model, the same test produces a *P*-value of 0.2380 in assortativity and 0.9522 in Clustering Coefficient; there is no evidence to reject our model. Results for the other sampling techniques can be found in Supplementary Material S6.

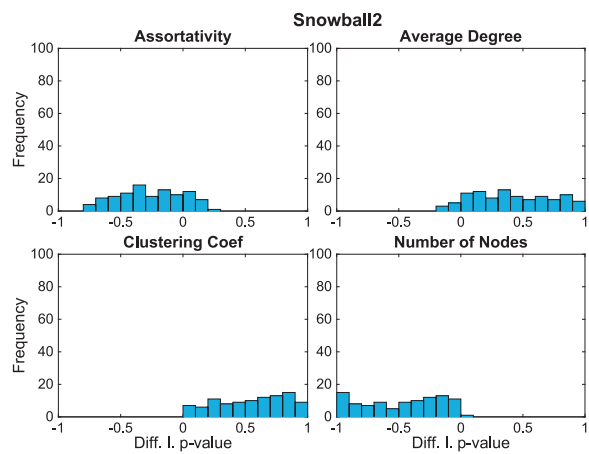

**Fig. 4.** Histogram of differences in *P*-values of 100 2-hop Snowball Sample in the BioGRID PPI network with 25 initial random seed proteins and a bin size of 20 generated by adding additional redundant seed nodes. Each of the *P*-values are computed using 2000 Monte Carlo realizations (Color version of this figure is available at Bioinformatics online.)

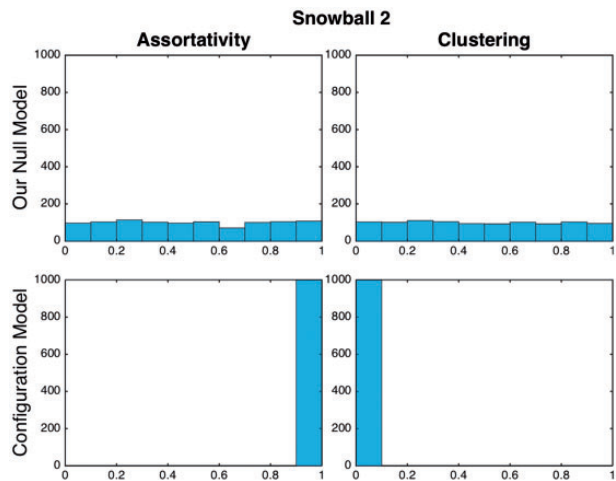

**Fig. 5**. Distribution of *P*-value results for 2-hop snowball sampling under our null model and the Configuration Model. 1000 networks are generated by selecting 25 random seeds, assortativity and average local clustering coefficient are calculated. For each network we calculate the *P*-value with respect to a random network (under our null model and the configuration model) (Color version of this figure is available at Bioinformatics online.)

While we cannot generalize from these results to all possible networks ensembles, and it is highly likely that there are network models and parameters ranges where the configuration model performs well in subnetworks, the configuration model does not perform well in general when comparing subnetworks based on seed lists. This demonstrates the need for an alternative to the configuration model for this task.

## 4 Discussion

There is a need for a robust and reliable nonparametric test when testing the significance of summary statistics for sampling techniques based on seed lists. Depending on the research question the configuration model does not fulfil this role. Here we propose an alternative null model that is based on an ensemble of seed lists generated from the minimum seed list.

We focus on the significance of network features, given a construction method, rather than given a construction method and a fixed seed list, as different seed lists may result in the same subnetwork.

We have demonstrated that accounting for seed list construction is important, by artificially increasing the significance of a randomly chosen seed lists in a biological network, and through observing the effect of this increase on the biologically motivated seed lists.

We have also shown through our benchmark that *P*-values from our test are negatively correlated in all but one case with measures of purity and accuracy of the sample (i.e. on average small *P*-values result in more accurate/pure networks).

Our null model is not without issues. Notably, it is rare but possible for there to be more than one minimum seed list which then requires a comparison with multiple seed lists. A further problem is that the seed list does not have to be a global minimum; it is possible that there is a seed list that is smaller than the supposed 'minimum seed list'. Finding this minimum seed list for an arbitrary technique is computationally prohibitive. We believe that the very tractable null model presented in this paper is superior to the model based on a globally minimum seed list, due to its applicability to many different problems.

For PPI networks, our nonparametric test allows us to choose a subnetwork which may have interesting properties for further analysis for our work on PD. On the statistics tested many of the generated networks do not appear to deviate significantly from random, unlike the results from the Expression seed list using Path2 and Shortest Paths. Our work also highlights the need to focus more attention on generative models of biological networks in order to generate parametric models of these systems.

## Acknowledgements

We would like to thank Harriet A. Keane for very helpful conversations on Parkinson's disease. We would also like to thank Griffith Rees, Charlotte Deane and Marta Sarzynska for many helpful discussions.

## Funding

We acknowledge e-Therapeutics plc and the UK's Engineering and Physical Sciences Research Council (EPSRC) for funding, via a studentship at the Systems Approaches to Biomedical Science Industrial Doctorate Centre at the University of Oxford. GR acknowledges support from EPSRC grant EP/K032402/1 and from the Oxford Martin School programme on Resource Stewardship. FRT acknowledges support from James Martin 21st Century Foundation grant LC1213-006.

*Conflict of Interest*: none declared.

# Supplementary Information: A Nonparametric Significance Test for Sampled Networks

Andrew Elliott[1,*], Elizabeth Leicht[1], Alan Whitmore[4], Gesine Reinert[2], and Felix Reed-Tsochas[1,3]

[1]CABDyN Complexity Centre, Saïd Business School, University of Oxford
[2]Department of Statistics, University of Oxford
[3]Oxford Martin School, University of Oxford
[4]e-Therapeutics plc, 17 Blenheim Office Park, Long Hanborough, OX29 8LN
[*]To whom correspondence should be addressed.

July 5, 2017

# 1 Seed List Construction

This section contains the tables with the seed lists and the conversions into BioGRID identifiers. Note, some of the identifiers convert to more than one node in BioGRID, and some are not present.

The seeds for OMIM can be found in Table 1. In the case of the table for the expression seed list as this article is open access, copyright issues prevent us from reprinting the original table from Ref. [2], thus we include the conversions we made from the table into gene names and from these gene names to BioGRID IDs (Table 2) and we invite the reader to observe Table 1 in Ref. [2].

Note that when constructing this list we were unable successfully to map two of the genes in the expression seed into gene names. Subsequently we have been able to do so and this will be incorporated into future work. However, the results in this document are based on the list without these genes, as is reflected in the table.

| OMIM Entry | Mapped BioGrid Ids |
|---|---|
| Parkinson disease 1, 168601 (3)—SNCA, NACP, PARK1, PARK4—163890—4q22.1 | 112506 |
| Parkinson disease 10 (2)—PARK10, AAOPD—606852—1p32 | NONE |
| Parkinson disease 11, 607688 (3)—GIGYF2, KIAA0642, PARK11—612003—2q37.1 | 117520 |
| Parkinson disease 12 (2)—PARK12—300557—Xq21-q25 | NONE |
| Parkinson disease 13, 610297 (3)—HTRA2, OMI, PARK13, PRSS25—606441—2p13.1 | 118165 |
| Parkinson disease 14, 612953 (3)—PLA2G6, IPLA2, INAD1, NBIA2B, NBIA2A, PARK14—603604—22q13.1 | 113986* |
| Parkinson disease 15, autosomal recessive, 260300 (3)—FBXO7, FBX7, FBX, PKPS, PARK15—605648—22q12.3 | 117326 |
| Parkinson disease 17, 614203 (3)—VPS35, MEM3, PARK17—601501—16q11.2 | 120855 |
| Parkinson disease 18, 614251 (3)—EIF4G1, EIF4G, PARK18—600495—3q27.1 | 108296 |
| Parkinson disease 3 (2)—PARK3—602404—2p13 | NONE |
| Parkinson disease 4, 605543 (3)—SNCA, NACP, PARK1, PARK4—163890—4q22.1 | 112506 |
| Parkinson disease 6, early onset, 605909 (3)—PINK1, PARK6—608309—1p36.12 | 122376 |
| Parkinson disease 7, autosomal recessive early-onset, 606324 (3)—DJ1, PARK7—602533—1p36.23 | 116446 |
| Parkinson disease 8, 607060 (3)—LRRK2, PARK8—609007—12q12 | 125700 |
| Parkinson disease 9, 606693 (3)—ATP13A2, PARK9, KRPPD—610513—1p36.13 | 116973 |
| Parkinson disease, juvenile, type 2, 600116 (3)—PRKN, PARK2, PDJ, LPRS2—602544—6q26 | 111105 |
| Parkinson disease 16 (2)—PARK16—613164—1q32 | NONE |
| Parkinson disease 5, susceptibility to, 613643 (3)—UCHL1, PARK5—191342—4p13 | 113192 |
| Parkinson disease, late-onset, susceptibility to, 168600 (3)—GBA—606463—1q22 | 108899* |
| Parkinson disease, susceptibility to, 168600 (3)—ADH1C, ADH3—103730—4q23 | NONE |
| Parkinson disease, susceptibility to, 168600 (3)—MAPT, MTBT1, DDPAC, MSTD—157140—17q21.31 | 110308* |
| Parkinson disease, susceptibility to, 168600 (3)—TBP, SCA17, HDL4—600075—6q27 | 112771 |

Table 1: OMIM Seed List including mappings to BioGRID. OMIM records are taken from the OMIM database download [5] (full citation is as follows: Online Mendelian Inheritance in Man, OMIM. McKusick-Nathans Institute of Genetic Medicine, Johns Hopkins University (Baltimore, MD), 2012. World Wide Web URL: https://omim.org/) Conversion is performed using the mapping present in the BioGRID database. Proteins marked with a * are present in BioGrid but are not part of the largest connected component of the network composed of interactions from Yeast 2 Hybrid.

| No. | Gene Name | Mapped BioGrid Ids |
|---|---|---|
| 1 | LGALS3 | 110149 |
| 2 | SOX4 | 112542 |
| 3 | TWIST1 | 113142 |
| 4 | PGK1 | 111251 |
| 5 | NOT IDENTIFIED | NONE |
| 6 | SNRPB2 | 112513 |
| 7 | CLK3 | 107609 |
| 8 | CDKN1A | 107460, 111099 |
| 9 | NFE2L1 | 110851 |
| 10 | ASNS | 106932 |
| 11 | ID3 | 109625 |
| 12 | ID1 | 109623 |
| 13 | PTMA | 111724, 111728* |
| 14 | IGFBP5 | 109709 |
| 15 | JUN | 109928 |
| 16 | MYC | 110694 |
| 17 | RBM14 | 115700 |
| 18 | NR2F1 | 112883, 112884 |
| 19 | HSPA1A | 109535, 109536 |
| 20 | HSPA9 | 109545 |
| 21 | SLC12A4 | 112449* |
| 22 | COMT | 107707 |
| 23 | PLOD1 | 111366 |
| 24 | GOT1 | 109067, 119232* |
| 25 | MTHFD2 | 116011* |
| 26 | CCT5 | 116603 |
| 27 | RPN2 | 112100 |
| 28 | GADD45A | 108014 |
| 29 | GSTT2 | NONE |
| 30 | DDIT3 | 108016 |
| 31 | SIAH2 | 112373 |
| 32 | SSRP1 | 112627 |
| 33 | IL2RA | 109774* |
| 34 | DLK1 | 114316 |
| 35 | GPI | 115325, 109082* |
| 36 | VEGFA | 113265 |
| 37 | RNH1 | 111977*, 128882, 129787* |
| 38 | NOT IDENTIFIED | NONE |
| 39 | PRKCA | 111564 |
| 40 | PPP2R2B | 111513 |
| 41 | RAC1 | 111817 |
| 42 | YWHAB | 113361 |
| 43 | CTTN | 108332 |
| 44 | HINT1 | 109341 |
| 45 | SOCS2 | 114362 |
| 46 | CRYM | 107815 |
| 47 | SOCS7 | 119053, 125796 |
| 48 | TSC2 | 113100 |

Table 2: Table contains our conversions of the genes in Table 1 from Ref [2] into gene names and the subsequent conversion into a BioGRID identifier using the BioGRIDs internal conversions. The ordering is the same in each table. Proteins marked with a * are present in BioGRID but are not part of the largest connected component of the network composed of interactions from Yeast 2 Hybrid.

## 2 Derivation Of Analytic Results

The analytic results shown here have been extended from results in Ref. [4] and follow some of the notation. Note, Ref. [4] focuses on the classic problem of an unknown network with observed samples, in this case we have the related problem where we know that the network but we want to explore features of the samples. We define the function $h(J, s)$ as the probability that there does not exist a node within $n$ hops of $J$ on a randomly selected seed list of size $s$. Let $B_n(J)$ be the set of nodes within $n$ hops of the nodes in $J$. If the selection is uniformly at random from all subsets of nodes of size $s$ then we can use a hypergeometric distribution to derive an expression for $h(J, s)$. We can do so as follows, we randomly select $s$ seeds from $|V|$, however if we select a seed from $B_n(J)$ then at least one node in $J$ will be included. Thus we require the seeds to all be chosen from $V \setminus B_n(J)$. Thus through the hypergeometric distribution this results in the following form for $h(J, s)$:

$$h(J,s) = \frac{\binom{|V|-|B_n(J)|}{s}\binom{|V|-(|V|-|B_n(J)|)}{s-s}}{\binom{|V|}{s}},$$

which simplifies to:

$$h(J,s) = \frac{(|V|-s)!(|V|-|B_n(J)|)!}{(|V|-|B_n(J)|-s)!|V|!} = \prod_{i=0}^{s-1} \frac{|V|-|B_n(J)|-i}{|V|-i},$$

If we wish to fix the degree sequence of the seeds (or with a small adjustment binned degree), we can use a similar approach to the uniform case, to derive an expression for a degree sequence version, $h_d(J,t)$ where $t$ is the degree sequence. This results in the following form for $h(J,t)$:

$$h_d(J,t) = \prod_{u \in U(t)} \prod_{i=0}^{F(t,u)-1} \frac{D(V,u) - D(B_n(J),u) - i}{D(V,u) - i},$$

where $t$ is the degree sequence of the seed list. Here $F(t,u)$ is a counting function, it counts how many instances of $u$ there are in $t$ (e.g. $F([1,2,3,4,1],1) = 2$), $U(t)$ returns the unique elements of $l$ and $D(J,r)$ is the number of elements in $J$ of degree $r$. The seeds of different degrees are selected independently so we can calculate the probability for each unique degree in the seed degree list and then multiply them to get the final probability.

**Deriving the Mean and Variance** Following the notation in the paper we let $X$ be a random variable denoting the number of nodes in a snowball sampled graph with seed list $S$, where $S$ is a uniform random draw over all possible seed lists. For notational convenience we define $|S|$ as the number of seeds in $S$. We are interested in the case where $|S| = s$, thus we will constrain our calculations to this case. Note, by further restricting $S$ we obtain other schemes for example enforcing the degree sequence. In the case of enforcing the degree sequence we can repeat many of the following arguments by replacing $h$ with $h_d$.

If we let $Y_i$ be an indicator variable for the presence of node $i$ in the sample, then the number of nodes in a sample is $X = \sum_i Y_i$. We can compute the mean number of nodes as follows:

$$E\big[X\big|\ |S|=s\big] = E\Big[\sum_i Y_i\Big|\ |S|=s\Big] = \sum_i E\big[Y_i\big|\ |S|=s\big] = \sum_i 1 - h(\{i\},s) = |V| - \sum_i h(\{i\},s).$$

To compute the variance we can use the correlated variables formula to get:

$$\text{Var}(\sum_i Y_i\big|\ |S|=s) = \sum_i \text{Var}(Y_i\big|\ |S|=s) + 2\sum_{i<j} \text{Cov}(Y_i,Y_j\big|\ |S|=s).$$

We can imagine each term in the summation $(Y_i|\ |S|=s)$ as a Bernoulli random variable with $p = 1 - h(\{i\},s)$, therefore,

$$\text{Var}(Y_i\big|\ |S|=s) = p(1-p) = h(\{i\},s)(1-h(\{i\},s)).$$

The covariance between the two variables is defined as:

$$\text{Cov}(Y_i,Y_j\big|\ |S|=s) = E\big[Y_iY_j\big|\ |S|=s\big] - E\big[Y_i\big|\ |S|=s\big]E\big[Y_j\big|\ |S|=s\big];$$

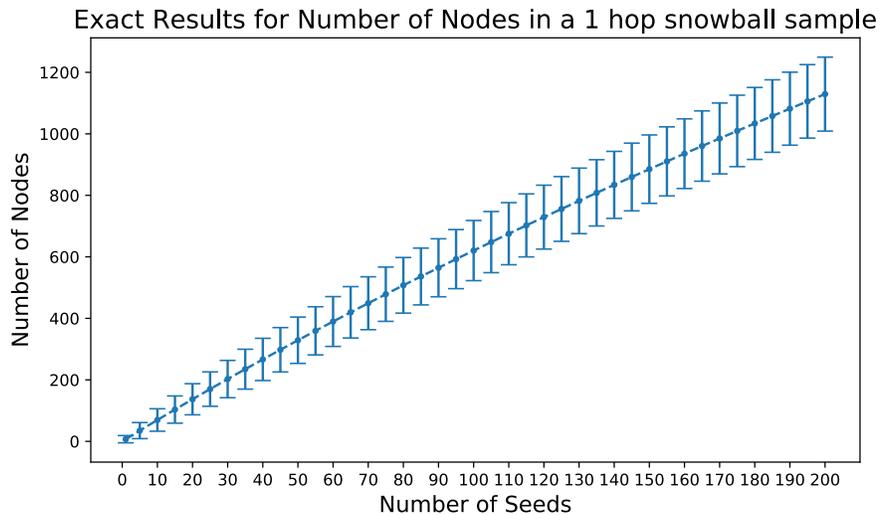

Figure 1: Exact Results: Points represent the mean number of nodes given a number of seed nodes in a 1-hop snowball sample from the BioGRID PPI network. The error bars represent one standard deviation of the same quantity.

thus:
$$E\big[Y_iY_j\big|\,|S|=s\big] = E\big[(1-(1-Y_i))(1-(1-Y_j))\big|\,|S|=s\big]$$
$$= 1 - E\big[1-Y_i\big|\,|S|=s\big] - E\big[1-Y_j\big|\,|S|=s\big] + E\big[(1-Y_i)(1-Y_j)\big|\,|S|=s\big]$$
$$= 1 - h(\{i\},s) - h(\{j\},s) + h(\{i,j\},s).$$

Combining and simplifying all of the terms we obtain:
$$\mathrm{Var}\Big(\sum_i Y_i\Big|\,|S|=s\Big) = \Big(\sum_i h(\{i\},s) - h(\{i\},s)^2\Big) + 2\Big(\sum_{i<j} h(\{i,j\},s) - h(\{i\},s)h(\{j\},s)\Big).$$

We can then factor the expression to obtain the form in the paper:
$$\mathrm{Var}\Big(\sum_i Y_i\Big|\,|S|=s\Big) = \sum_{\alpha=1}^{2} \alpha \sum_{\substack{J\subseteq V \\ |J|=\alpha}} h(J,s) - \Big(\sum_i h(\{i\},s)\Big)^2.$$

where $L$ is a dummy summation variable.

Figure 1 illustrates the effect of seed list size on the distribution of the number of nodes in a 1-hop snowball sample in the BioGRID PPI network [1, 7]. As we do not have a seed list of interest in this case, we have assumed that there is no restriction on the degree sequence of the seed nodes.

As expected, the larger the number of seed nodes, the larger the average number of nodes in the resulting network. Further, a small change in the number of seed nodes can have a large impact on the expected size of the network. The number of nodes in a 1-hop snowball sample on, say 20 proteins from the PPI network may appear small when compared to subnetworks randomly generated from 30 seeds but large when compared to such subnetwork generated from 10 seeds.

## 3 Algorithms To Add Or Remove Redundant Seeds

To discover redundant seed nodes we need to be able to guarantee that a pair of labelled networks are identical. The trivial way to do this is to compare the node lists and the edge lists, and if they are equal then the networks are equal. As we will have to check equality a large number of times, this approach can can prove computationally constraining. Making use of the fact that we are adding or removing nodes from the seed list, we can derive a simpler condition for this problem.

We take an arbitrary network with node set $V$ and edge list $E$ and a seed list $l_1$. Let $[V_{l_1}, E_{l_1}] = f_{arb}(V, E, l_1)$, where $f_{arb}$ is a arbitrary network sampling function, $l_1$ is a seed list and $V_{l_1}$ and $E_{l_1}$ are the subset of nodes and edges that are in the subnetwork.

Let us assume that we have two seed lists $l_1$ and $l_2$ and $l_1 \subseteq l_2$. Further, let us assume that our sampling technique ($f_{arb}$) guarantees that $V_{l_1} \subseteq V_{l_2}$ and $E_{l_1} \subseteq E_{l_2}$ if $l_1 \subseteq l_2$.

Trivially for sets $T_1$ and $T_2$ if $T_1 \subseteq T_2$ and $|T_1| = |T_2|$, then $T_1 = T_2$. Thus, if $|E_{l_1}| = |E_{l_2}|$ and $|V_{l_1}| = |V_{l_2}|$ then $E_{l_1} = E_{l_2}$ and $V_{l_1} = V_{l_2}$. Therefore we can use the condition:

$$|E_{l_1}| = |E_{l_2}| \text{ and } |V_{l_1}| = |V_{l_2}|. \tag{1}$$

Note, we must condition both on the set of edges and the set of nodes as they are both required to fully define the subnetwork.

Therefore if we can guarantee using our sampling techniques that if $l_1 \subseteq l_2$ then $V_{l_1} \subseteq V_{l_2}$ and $E_{l_1} \subseteq E_{l_2}$ then we can simply test for equality in the number of nodes and edges.

**Snowball Sampling** In snowball sampling the contribution from each of the nodes on the seed list is independent, as it is simply the number of nodes within a certain radius of the each of the seeds. Therefore the expression for the $V_{l_1}$ is as follows:

$$V_{l_1} = \bigcup_{x \in l_1} V_x.$$

If we take $l_1 \subseteq l_2$, then $V_{l_2} = V_{l_1} \cup V_{l_2 \setminus l_1}$ and therefore trivially $V_{l_1} \subseteq V_{l_2}$. We can use the same argument for $E_{l_1}$. We can therefore use the condition in (1) for snowball sampled networks given that the seed list is a subset or a superset of the original seed list.

**Deterministic Path Based Sampling Techniques** We define a deterministic path based sampling techniques in undirected networks as a procedure for which the sample as is function of every pair of nodes on the seed list and the underlying network. Note that Path≤k and shortest path sampling both fall into this category. Therefore,

$$V_{l_1} = \{x \in P_{arb}^{(V)}(y, z) : y, z \in l_1\}, \text{ and } E_{l_1} = \{x \in P_{arb}^{(E)}(y, z) : y, z \in l_1\},$$

where $P_{arb}^{(V)}(y, z)$ and $P_{arb}^{(E)}(y, z)$ are sets of the sampled nodes and edges respectively on the path(s) between $y$ and $z$ defined by the sampling technique in question. Let $l_1$ and $l_2$ be seed lists and let $l_1 \subseteq l_2$, then

$$V_{l_2} = V_{l_1} \cup V_{l_2 \setminus l_1} \cup \{x \in P_{arb}^{(V)}(y, z) : y \in l_1, \ z \in l_2 \setminus l_1\} \cup \{x \in P_{arb}^{(V)}(y, z) : z \in l_2, \ y \in l_2 \setminus l_1\}.$$

where the right hand side is the union of the nodes included by the seed list $l_1$ (first term), the nodes that are included by the seed list $l_2 \setminus l_1$ (second term), and the third and fourth terms represent the contributions from paths which connect a node from $l_1$ to a node from $l_2 \setminus l_1$. Therefore $V_{l_1} \subseteq V_{l_2}$, and by similar argument $E_{l_1} \subseteq E_{l_2}$. Thus we can use the condition in (1) for all deterministic path sampling techniques.

## 3.1 Algorithm To Add Additional Redundant Seed Nodes

In the paper given a seed list, a network, a sampling technique and the resultant subnetwork, we construct the largest seed list that will generate the same subnetwork. For a given seed list and network the procedure to do this is as follows:

1. Make an empty list, which will contain the possible seed nodes.

2. For each node in the subnetwork, compute if it can be added to the seed list without increasing the number of nodes or edges in the subnetwork. If so add to the list of possible seed nodes.

3. Set $W = 0$.

4. Form all seed lists with the original seeds and all but $W$ of the nodes on the possible seed list.

5. Test if each of these seed lists produces the same subnetwork. If at least one seed list produces the subnetwork return all tested seed lists that produce the same subnetwork, else let $W = W + 1$ and go back to step 4.

For all results in the paper we used the algorithm above, however for some sampling techniques we can construct a simplified procedure. For $n$-hop snowball sampling, we can construct a list of possible seed nodes using the following procedure:

```
listofSeeds=[]
OutsideNodes=list of nodes in (n+1)-hop snowball sample but not in n-hop sample
for curNode in Subnetwork
    if a node in OutsideNodes is in the n-hop snowball sample of curNode:
        continue
    else:
        add curNode to listofSeeds
```

## 3.2 Optimising Finding Redundant Seed Nodes

The algorithm presented in Section 2.4 of the paper is the following:

1. Remove each seed in turn and check if the number of nodes and edges in the subnetwork do not change. If not, then add the node to the list of redundant seeds.

2. Form a list of the remaining seeds.

3. Define a dummy variable L and set $L = 0$

4. For lists of redundant seeds of length $L$

    (a) Test if sampling with the list of the remaining seeds and the selected redundant seeds produce the same network.

    (b) Store all seed lists which pass the test.

5. If there are no seed lists which pass the test, set $L \to L + 1$ and go to step 4.

6. Return the smallest seed list(s) that produce the same network.

The major problem in this procedure is the large number of options that may need to be checked to find the minimum seed list. As stated in the paper finding the minimum seed list for snowball sampled networks can be converted into the set cover problem which is NP-hard [3].

The set cover problem is defined as follows, for a set $\mathfrak{F}$, and a collection of subsets $F = \{F_1, ..., F_m\}$ such that $\mathfrak{F} = \bigcup_{x \in F} x$ [3]. The problem is then to find the smallest subset of $F$ which we shall call $F^*$ such that $\mathfrak{F} = \bigcup_{x \in F^*} x$ [3].

We can reformulate the minimum seed list for Snowball sampled networks as follows. We let $F$ be an empty set. For each seed node we compute the set of nodes which are sampled by this seed node and we add it to $F$. Finding the minimum seed list is then equivalent to finding $F^*$ and then returning the seeds which were used to construct each element of $F^*$.

Thus in cases where we require additional speed we could try reformulating this as a set cover problem and use state of the art algorithms for this problem.

It may also be possible to convert the path based techniques into another related NP Complete or NP Hard problem and use a similar technique. However, as we do not require the speed for the work we are doing here we have not attempted to do so.

**Further Optimisations** A further optimisation, which is sampling technique dependent, can be performed on sampling techniques that scale with number of nodes in the whole network and that only depend on the information in the subnetwork. Sampling in the subnetwork rather than in the wider network can be more efficient while still guaranteeing the result. One example of this procedure is shortest path sampling. All of the information about shortest paths is included in the subnetwork. Therefore sampling with the reduced seed list in the subnetwork saves time (as shortest path scales with number of nodes and edges depending on implementation) and guarantees that the result is correct as long as the seed list is a subset of the original seed list.

# 4 Further Results: Adding Redundant Seed Results

We showed that redundant seed nodes have to be taken into account; in particular in Figure 4 of the paper we demonstrated that the significance of randomly chosen seed lists can be changed in the BioGRID network under 2-hop snowball sampling by increasing the size of the seed list without changing the resultant sampled network. A similar effect can be observed several other sampling techniques as can be seen in Figure 2.

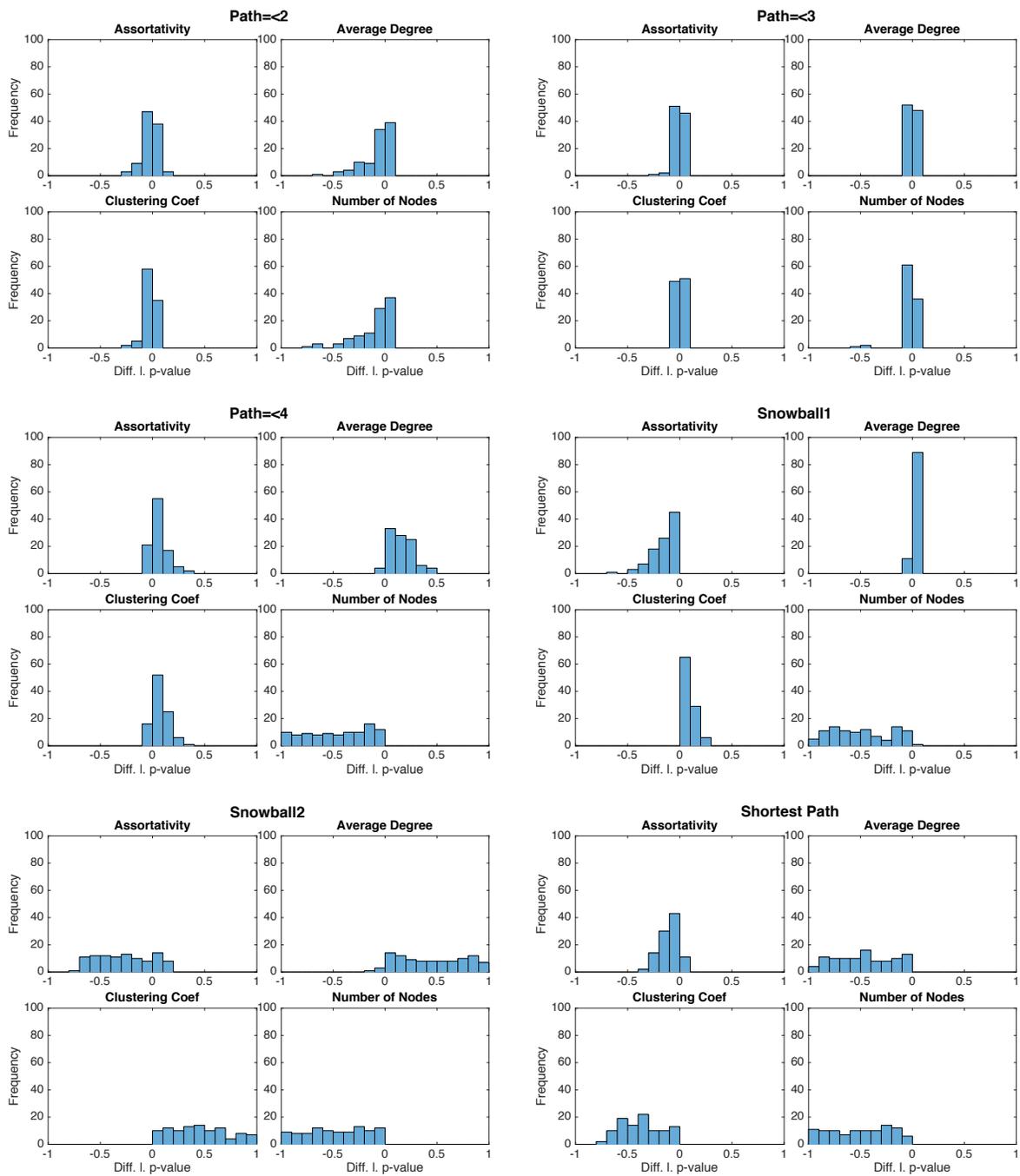

Figure 2: Histogram of differences in $p$-values of 100 of each of the sampling techniques in the BioGRID PPI network with 25 initial random seed proteins and a bin size of 20 generated by adding additional redundant seed nodes. Each of the $p$-values are computed using 2000 Monte Carlo realisations. In several cases we observe a large change in $p$-value using this procedure.

# 5 Empirical Seed Lists: Additional Results

To test whether the results we see in the paper are robust with respect to the bins that we use to generate the random seed lists, we recalculate the results in the paper using a minimum bin size of 5, 10, 20, 30 and 50. The smaller the bin size the closer the degree sequence will match the test sequence, whereas the larger minimum bin sizes produce seed lists which have degree sequences which are further from the test list, but have a lower likelihood of selecting the same small set of seed nodes. Figure 3 shows the results for the OMIM seed list and Figure 4 shows the results for the expression seed list.

# 6 Comparison With Configuration Model

The configuration model may not preserve the structural features of the original network. For example in the 2-hop snowball sample in Figure 5 there is a very clear structure, with a maximum path length of 4 between any pair of nodes. This structure will not be preserved in the configuration model Fig. 6 shows a simple comparison between the distribution of shortest path length in the snowball sampled network compared against an ensemble of configuration models of the same network.

In the paper we examine the $p$-value distribution using our null model and the configuration model in 2-hop snowball sampled subnetworks. We repeated the comparison with all of the other sampling techniques using the method described in the main paper. The results can be seen in Figure 7. We use a $\chi^2$ test to compare the distributions with the uniform distribution taking as observations the $p$-values of the statistic of interest from 1000 networks generated by selecting 25 random seeds (Table 3). We see that in all sampling techniques we reject the null hypothesis for the configuration model and for Snow1, Snow2 and all shortest paths we do not reject the null hypothesis for our null model. For Path3 in clustering we reject the null hypothesis at the 5% level but not at the 1% and further visual inspection of the distribution also does not draw any concern.

In the case of the clustering in Path2, triangles can occur only when two seed nodes are less than or equal to 2 hops apart and one of them is part of a triangle with $8,292$ nodes and an average shortest path length of 4.35 this is unlikely to happen. For non-continuous distributions, rather than the uniform distribution, we would expect to see the generalised inverse of the cumulative density function as distribution of the $p$-values. Fig. 8 shows that the distribution of average local clustering coefficients for Path2 networks sampled with 25 random seeds is indeed discontinuous and therefore we should not be surprised to see a non-uniform null distribution. In contrast with the same distribution from Snow1 shown in Fig. 9 is approximately continuously distributed and therefore the uniform distribution appears as the null distribution as expected.

While we cannot generalise from these results to all possible networks ensembles, and it is highly likely that there are network models and parameters ranges where the configuration model performs well in subnetworks, the configuration model does not perform well in general when comparing subnetworks based on seed lists. This demonstrates the need for an alternative to the configuration model for this task.

|          | Our Null       |            | Configuration model |            |
|----------|----------------|------------|---------------------|------------|
|          | Assortativity  | Clustering | Assortativity       | Clustering |
| Path3    | 0.3115         | 0.0255     | 0                   | 0          |
| Path4    | 0.9659         | 0.0734     | 0                   | 0          |
| Shortest | 0.8343         | 0.4788     | 0                   | 0          |
| Snow1    | 0.4654         | 0.4788     | 0                   | 0          |
| Snow2    | 0.2380         | 0.9522     | 0                   | 0          |

Table 3: Goodness of fit $\chi^2$ $p$-value results for different sampling techniques under our null model and the configuration model. 1000 networks are generated by selecting 25 random seeds and sampling with the given technique. Assortativity and average local clustering coefficient are calculated on the resultant network. We measure the $p$-value of these statistics under our null model and the configuration model. If the null hypothesis is correct and the distribution under the null hypothesis is continuous then these $p$-values, viewed as random observations, should be approximately uniformly distributed on the interval $[0, 1]$. This table gives the results from a $\chi^2$ test for goodness to the uniform distribution. The results for the configuration model indicate that the distribution of $p$-values is far from uniform indicating that the configuration model is not a good null model , whereas in our model we do not reject the null hypothesis, lending support to its use as a null model.

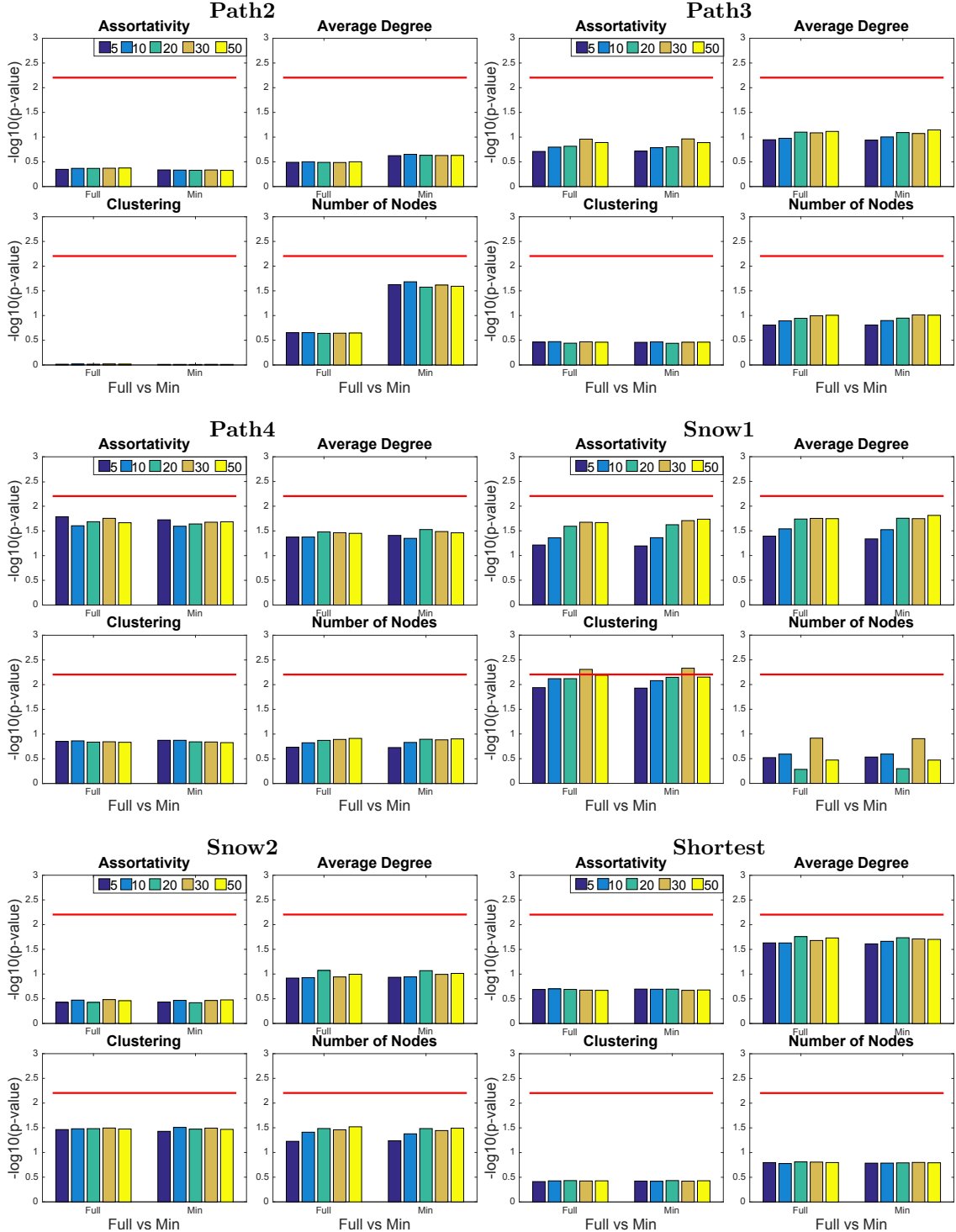

Figure 3: Test results for the OMIM seed list: smallest $p$-value, on a negative log scale under our null model. Multiple bars demonstrate the robustness of the method to the minimum number of items in each bin chosen when accounting for degree. Red: significance level (0.025/4). Left: Full seed list, right minimum seed list. Note negative logarithmic scale, e.g. only results that are above the Red significance line are considered significant. Only in Snowball 1 sampling is the result significant and only for the clustering coefficient, although this result is not robust to choice of bin size.

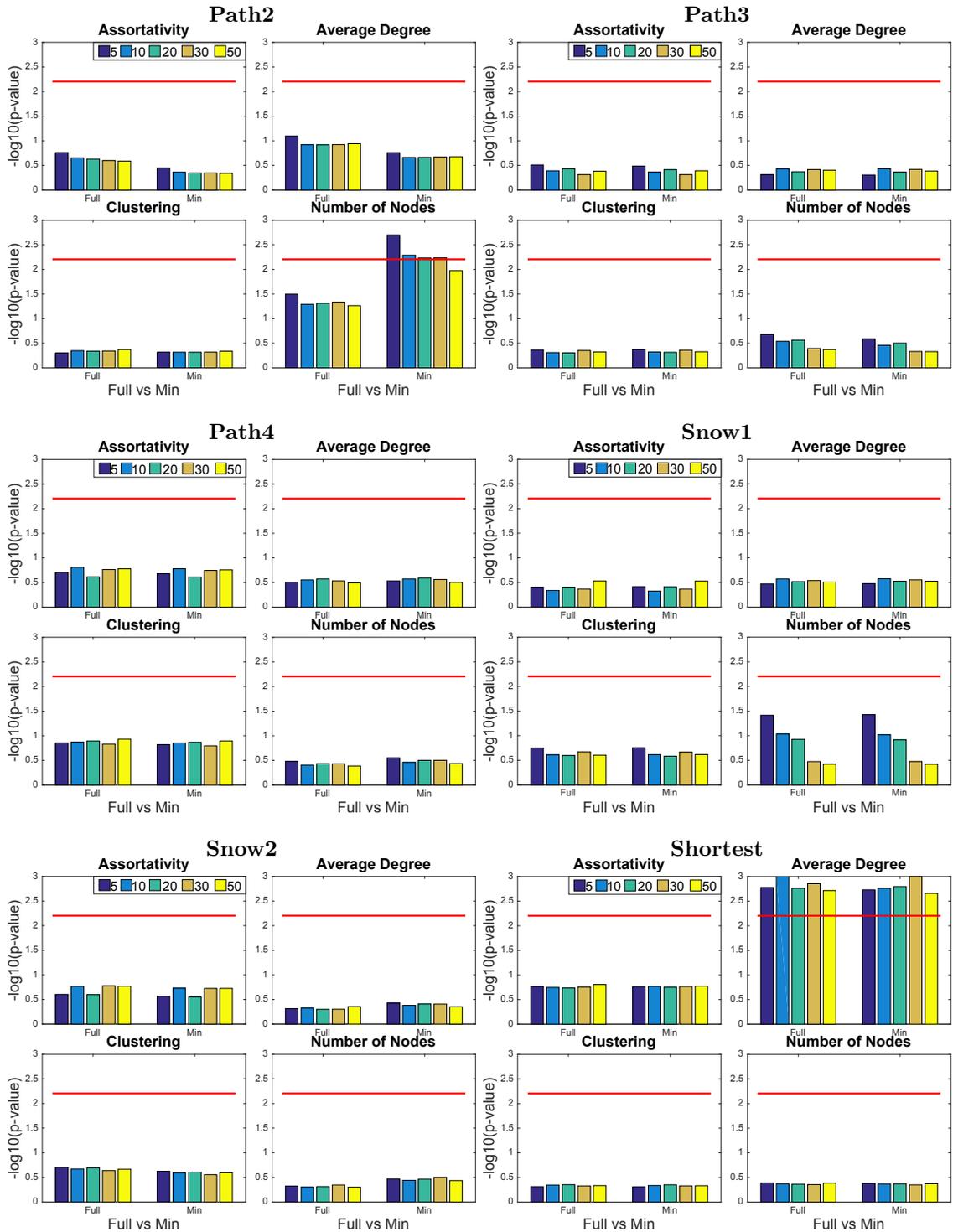

Figure 4: Test results for the Expression seed list: smallest $p$-value, on a negative log scale under our null model. Multiple bars demonstrate the robustness of the method to the minimum number of items in each bin chosen when accounting for degree. Red: significance level (0.025/4). Left: Full seed list, right minimum seed list. Note negative logarithmic scale, e.g. only results that are above the Red significance line are considered significant. Significant results are observed in number of nodes in Shortest path sampling and in most but not all bin sizes for the number of nodes in Path2 with the minimum seed list.

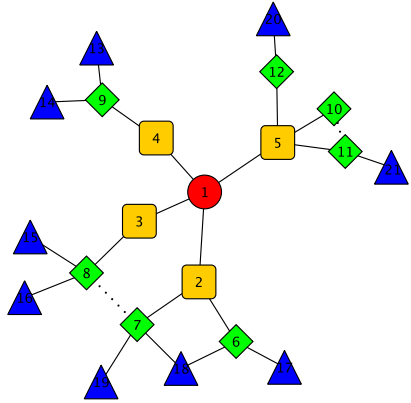

Figure 5: 2-Hop Snowball Sampling Example: The seed list consists of node 1 (circle), node shape represents distance from seed protein square, representing nodes 1 hop from the seed, diamond 2 hops from a seed, triangle 3 hops from a seed. Dashed edges represent cross-edges in a 2-hop snowball sample.

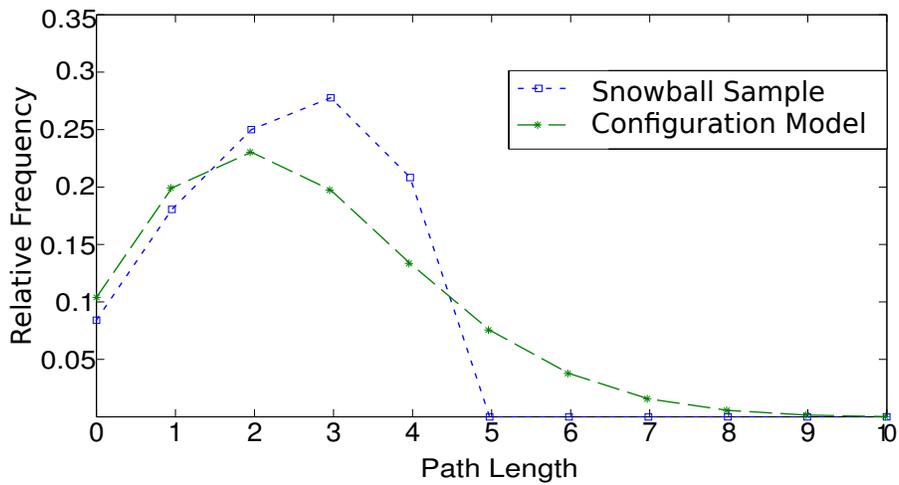

Figure 6: Comparison of distribution of shortest path lengths in the original network and over an ensemble of networks generated by a configuration model from the same network.
.

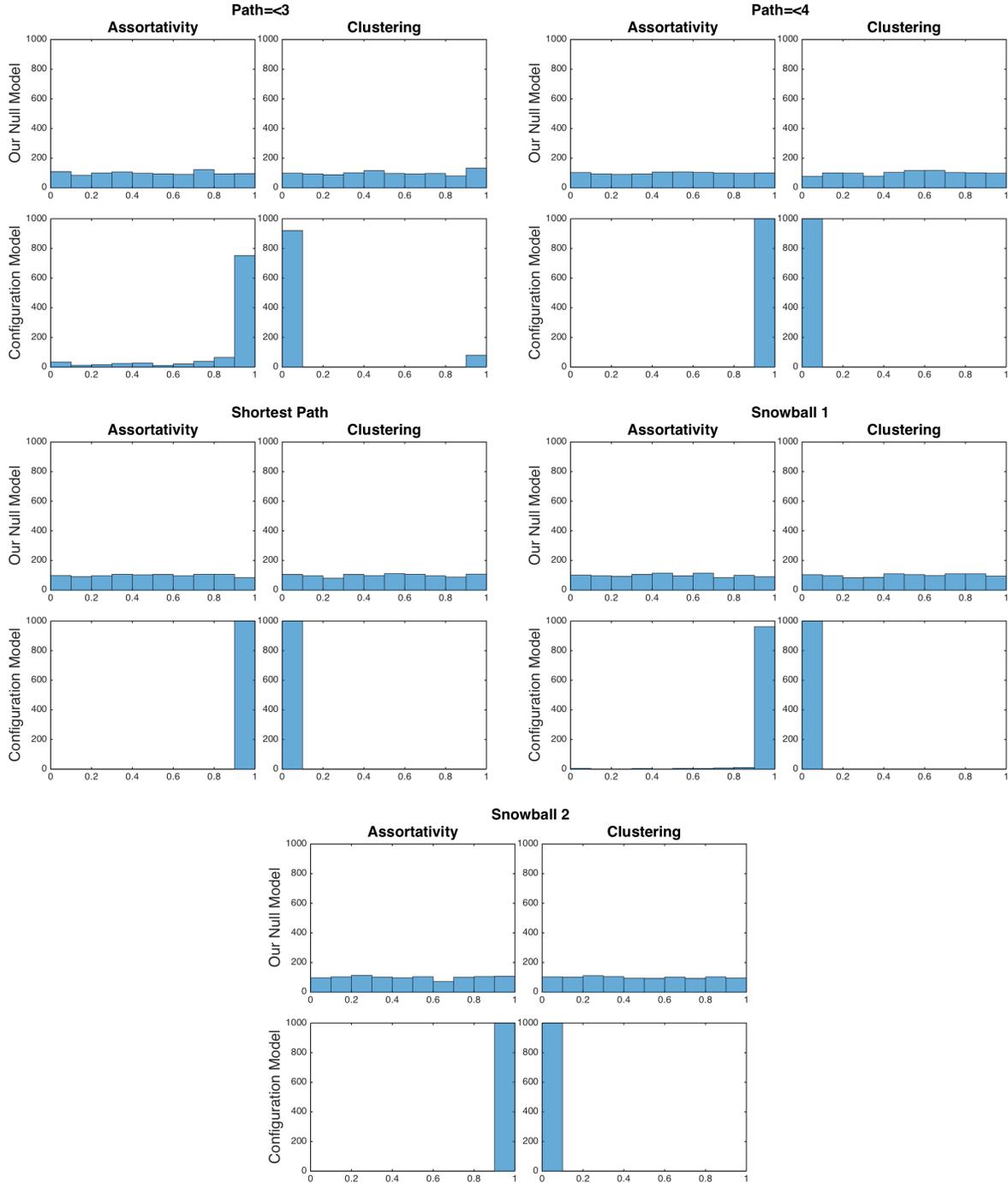

Figure 7: Distribution of $p$-value results for different sampling techniques under our null model and the Configuration Model. 1000 networks are generated by selecting 25 random seeds, assortativity and average local clustering coefficient are calculated. For each network we observe the $p$-value for deviation from a random network (our null model and the configuration model). For our null model the $p$-values are approximately uniformly distributed, indicating a reasonable fit of the null model. For the configuration model the distribution is far from uniform, indicating a poor fit.

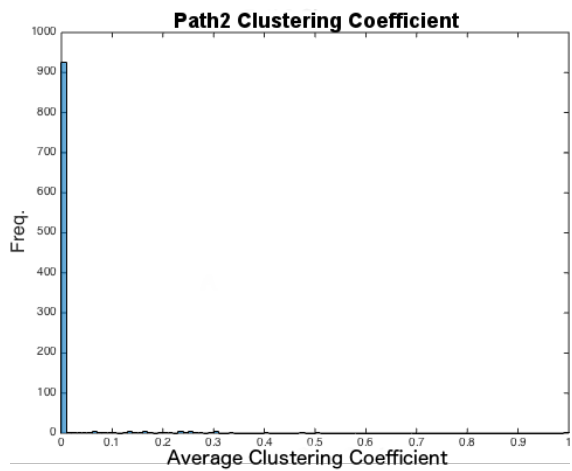

Figure 8: Distribution of Average Clustering Coefficient Path2 sampled from BioGRID network with 25 randomly chosen seeds. There is a strong concentration at 0 and a few discrete values which are not equal to 0.

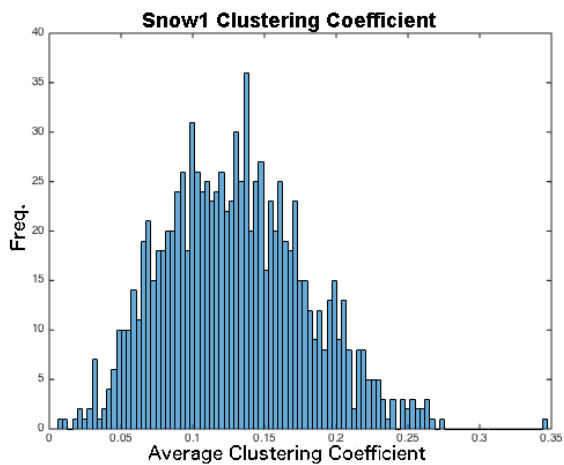

Figure 9: Distribution of Average Clustering Coefficient Snow1 sampled from BioGRID network with 25 randomly chosen seeds.